\documentclass{article}

\usepackage[final,nonatbib]{nips_2018}

\usepackage[nonatbib]{nips_2018}

\usepackage[utf8]{inputenc} 
\usepackage[T1]{fontenc}    
\usepackage{hyperref}       
\usepackage{url}            
\usepackage{booktabs}       
\usepackage{amsfonts}       
\usepackage{nicefrac}       
\usepackage{microtype}      
\usepackage{amsmath}
\usepackage{graphicx}
\usepackage{subcaption}
\usepackage{float}
\usepackage{caption}
\usepackage[compact]{titlesec}
\titlespacing{\section}{0pt}{*0}{*0}
\titlespacing{\subsection}{0pt}{*0}{*0}
\titlespacing{\subsubsection}{0pt}{*0}{*0}
\title{Reliable uncertainty estimate for antibiotic resistance classification with Stochastic Gradient Langevin Dynamics}

\author{
  Md-Nafiz Hamid\\
Bioinformatics and Computational Biology Program\\
 Department of Veterinary Microbiology and Preventive Medicine\\
  Iowa State University\\
  Ames, IA\\
  \texttt{nafizh@iastate.edu} \\
  \And
  Iddo Friedberg \\
 Bioinformatics and Computational Biology Program\\
 Department of Veterinary Microbiology and Preventive Medicine\\
  Iowa State University\\
  Ames, IA\\
  \texttt{idoerg@iastate.edu} \\
}

\begin{document}

\maketitle

\begin{abstract}
Antibiotic resistance monitoring is of paramount importance in the face of this ongoing global epidemic. Deep learning models trained with traditional optimization algorithms (e.g. Adam, SGD) provide poor posterior estimates when tested against out-of-distribution (OoD) antibiotic resistant/non-resistant genes. In this paper, we introduce a deep learning model trained with Stochastic Gradient Langevin Dynamics (SGLD) to classify antibiotic resistant genes. The model provides better uncertainty estimates when tested against OoD data compared to traditional optimization methods such as Adam. 

\end{abstract}

\section{Introduction}
Antibiotic resistance is a global scourge that is taking an increasing toll in mortality and morbidity, in both nosocomial and community acquired infections\cite{Neu1992Crisis,who2104amr}. A growing number of once easily treatable infectious diseases such as tuberculosis, gonorrhea, and pneumonia are becoming harder to treat as the scope of effective drugs is shrinking. The CDC estimates that 2,000,000 illnesses and 23,000 people die annually from antibiotic resistance in the US alone\cite{cdcamr}. The overuse of antibiotics in health care and agriculture is exacerbating the problem to the point that the World Health Organization is considering antibiotic resistance ``one of the biggest threats to global health, food security and human development today''. Identifying genes associated with antibiotic resistance  is an important first step towards dealing with the problem\cite{Brown2016Antibacterial}, and providing a narrow-spectrum treatment, targeted solely against the types of resistance displayed. This statement is especially true when dealing with genes acquired from human or environmental metagenomic samples\cite{Perry2014Antibiotic}. A rapid identification of the class of antibiotic resistance that may exist in a given environmental or clinical microbiome sample can provide immediate guidance to treatment and prevention.

In this study, we developed a deep neural network that can predict antibiotic resistance into 15 classes from protein sequences. It can be useful in identifying metagenomic sample resistance for the purpose of providing a focused drug treatment. Traditional methods \cite{kleinheinz2014applying, davis2016antimicrobial, pal2016structure} to identify antibiotic-resistant genes usually take a alignment based best-hit approach which causes the methods to produce many false negatives \cite{arango2018deeparg}. Recently, a deep learning based approach was developed that used normalized bit scores as features that were acquired after aligning against known antibiotic resistant genes \cite{arango2018deeparg}. In contrast, our model only uses the raw protein sequence as its input. At the same time, neural networks are known for providing high confidence scores on inputs that are from a different probability distribution than the model was trained on \cite{palacci2018scalable, choi2018generative}. This can result in disastrous consequences in sensitive applications such as health care or self-driving systems. Here, we develop two deep learning models that were trained with ADAM and SGLD. Both models give significant accuracy on the test set in terms of predicting antibiotic resistance solely from the protein sequence. But we show that the model trained with SGLD is better equipped to predict OoD data i.e., it assigns a low probability to sequences from proteins that are not related to antibiotic-resistance or are from classes that were not included in training. 

\section{Dataset and Model}
\label{model}
\paragraph{Dataset}
We used the dataset curated in the DeepArg study \cite{arango2018deeparg}. Briefly, The dataset was created from the CARD \cite{jia2016card}, ARDB \cite{liu2008ardb} and UNIPROT \cite{uniprot2018uniprot} databases with a combination of computational and manual curation. The original dataset has 14974 protein sequences that are resistant to 34 different antibiotics (our classes in the multi-class classification task). There were 19 classes that had training samples of 11 sequences or less. We discarded these classes and were left with 15 classes with a total of 14907 protein sequences.

\paragraph{Model} We used a self-attention based sentence embedding model introduced in \cite{Lin_Feng_Santos_Yu_Xiang_Zhou_Bengio_2017}. For input, we represented each amino acid in a protein sequence as a size 10 embedding that was randomly initialized, and then trained end-to-end. We used one single layer of LSTM with 64 units and a dropout value of 0.7. Following that is the self-attention part which we can think of as a feed-forward neural network with one hidden layer of 600 units. This network takes the output from the LSTM layer as input, and produces an output of size 100. We weighted this output with a softmax layer which outputs our attentions. We multiplied the outputs of the LSTM layer with these attentions to get a weighted view of the LSTM hidden states. The result of this multiplication became our sentence embedding for that specific protein sequence. 


\paragraph{Optimization}
Typically, neural networks are trained with optimization methods such as Stochastic gradient descent (SGD) \cite{robbins1985stochastic} or its variants such as Adam \cite{kingma2014adam}, Adagrad \cite{duchi2011adaptive}, RMSprop \cite{tieleman2012lecture} etc. In SGD, for each iteration a mini-batch from the dataset is used to update the parameters of the neural network. For each iteration $t$, training data $X_t = \{x_{t1},...,x_{tn}\}$ is provided, and for parameters $\theta$, the update $\Delta \theta_{t}$ is: 

\begin{equation}\label{sgd}
\Delta \theta_{t} = \frac{\epsilon_{t}}{2} \Bigg(\nabla \log p(\theta_{t}) + 
                         \frac{N}{n} \sum_{i=1}^{n} \nabla \log p(x_{ti}|\theta_{t})\Bigg)
\end{equation}

At the same time, SGD or its variants do not capture parameter uncertainty. In contrast, Bayesian approaches such as Markov Chain Monte Carlo (MCMC) \cite{robert2004monte} techniques do capture uncertainty estimates. One such class of techniques are Langevin dynamics \cite{roberts2002langevin} which inject Gaussian noise into Equation \ref{sgd} so that the parameters do not collapse into the Maximum a posteriori (MAP) solution:

\begin{align}\label{lang}
\Delta \theta_{t} = \frac{\epsilon}{2} \Bigg(\nabla \log p(\theta_{t}) + 
                         \frac{N}{n} \sum_{i=1}^{n} \nabla \log p(x_{ti}|\theta_{t})\Bigg) 
                         + \eta_{t}\\
where, \eta_{t} \sim N(0, \epsilon) \nonumber
\end{align}

However, MCMC techniques require that the algorithm go over the entire dataset per iteration before making a parameter update. This slows down the model training process, and also requires huge computational costs. To remove this problem, Stochastic Gradient Langevin Dynamics (SGLD) was introduced \cite{welling2011bayesian}, which combined the best of both worlds i.e. inserting Gaussian noise into each mini-batch of training data. In SGLD, during each iteration for SGD, Gaussian noise is injected which has a variance of the step-size $\epsilon_{t}$:
        
\begin{align}\label{lang}
\Delta \theta_{t} = \frac{\epsilon_{t}}{2} \Bigg(\nabla \log p(\theta_{t}) + 
                         \frac{N}{n} \sum_{i=1}^{n} \nabla \log p(x_{ti}|\theta_{t})\Bigg) 
                         + \eta_{t}\\
where, \eta_{t} \sim N(0, \epsilon_{t}) \nonumber
\end{align}

This injection of Gaussian noise has an advantageous side-effect, as it also provides a better calibration of confidence scores of predictions on OoD data. For example, \cite{palacci2018scalable} showed that an SGLD trained neural network provides low confidence scores when trained on the MNIST \cite{lecun2010mnist} dataset but tested on the NotMNIST dataset \cite{bulatov2011notmnist}; whereas an SGD trained neural network still naively provides high confidence scores.
We used SGLD to train a neural network to classify protein sequences into their antibiotic resistance classes. In the experiment section, we show that an SGLD trained network provides low confidence scores when predicting on OoD protein sequences while an ADAM trained model still provides high confidence scores.




\section{Experiment}
\label{experi}
The model that is trained with ADAM has the same self-attention architecture as the model used for SGLD training except it has 3 bi-directional LSTM layers. We used a learning rate of 0.001 with a weight decay value of 0.0001. 

We divided our dataset into a 70/20/10\% training, validation, and test set split. We trained our model with SGLD on the training dataset, and tuned the hyper-parameters by checking the performance on the validation dataset. Testing on the test dataset was done only once. 

Table \ref{comparison} shows the performance of both SGLD and ADAM trained models on the test set in terms of Precision, Recall and $F_1$ for each class and overall. We show that overall the ADAM trained model is performing better than the SGLD trained model. 

\begin{table}[]
\scriptsize
\centering
\caption{Comparison between SGLD and ADAM trained models for 15 different classes of antibiotic resistance.}
\label{comparison}
\begin{tabular}{|c|c|c|c|c|c|c|c|}
\hline
                                                                              & \multicolumn{3}{c|}{\textbf{SGLD trained model}}   & \multicolumn{3}{c|}{\textbf{ADAM trained model}}   &                                                                                      \\ \hline
\textbf{Antibiotics}                                                          & \textbf{Precision} & \textbf{Recall} & \textbf{F1} & \textbf{Precision} & \textbf{Recall} & \textbf{F1} & \textbf{\begin{tabular}[c]{@{}c@{}}Number of data points\\ in Test set\end{tabular}} \\ \hline
Multidrug                                                                     & 0.68               & 0.81            & 0.74        & 0.84               & 0.92            & 0.88        & 109                                                                                  \\ \hline
Beta Lactam                                                                   & 0.97               & 0.93            & 0.95        & 0.99               & 0.96            & 0.98        & 519                                                                                  \\ \hline
Aminoglycoside                                                                & 0.82               & 0.82            & 0.82        & 0.90               & 0.97            & 0.93        & 87                                                                                   \\ \hline
Rifampin                                                                      & 1.00               & 0.67            & 0.80        & 1.00               & 0.67            & 0.80        & 3                                                                                    \\ \hline
Tetracycline                                                                  & 0.68               & 0.70            & 0.69        & 0.86               & 0.70            & 0.78        & 27                                                                                   \\ \hline
Quinolone                                                                     & 0.75               & 0.92            & 0.83        & 0.80               & 0.92            & 0.86        & 13                                                                                   \\ \hline
\begin{tabular}[c]{@{}c@{}}Macrolide\\ lincosamide streptogramin\end{tabular} & 0.93               & 0.85            & 0.89        & 0.95               & 0.94            & 0.95        & 111                                                                                  \\ \hline
Fosfomycin                                                                    & 0.90               & 0.93            & 0.92        & 1.00               & 0.93            & 0.96        & 29                                                                                   \\ \hline
Polymyxin                                                                     & 0.97               & 0.97            & 0.97        & 1.00               & 0.99            & 0.99        & 90                                                                                   \\ \hline
Chloramphenicol                                                               & 0.78               & 0.83            & 0.80        & 0.98               & 0.89            & 0.93        & 47                                                                                   \\ \hline
Bacitracin                                                                    & 0.99               & 0.96            & 0.98        & 0.99               & 0.98            & 0.99        & 421                                                                                  \\ \hline
Kasugamycin                                                                   & 1.00               & 1.00            & 1.00        & 1.00               & 1.00            & 1.00        & 3                                                                                    \\ \hline
Trimethoprim                                                                  & 0.83               & 0.62            & 0.71        & 0.78               & 0.88            & 0.82        & 8                                                                                    \\ \hline
Sulfonamide                                                                   & 1.00               & 1.00            & 1.00        & 1.00               & 1.00            & 1.00        & 2                                                                                    \\ \hline
Glycopeptide                                                                  & 0.49               & 0.82            & 0.61        & 0.53               & 0.91            & 0.67        & 22                                                                                   \\ \hline
\textbf{Overall}                                                              & 0.92               & 0.91            & 0.91        & 0.96               & 0.95            & 0.96        &                                                                                      \\ \hline
\end{tabular}
\end{table}

In the next step, we tested both models on OoD samples. For this we used the 19 classes of antibiotic resistant classes we did not include in our dataset used for training and testing. These 19 classes have a total of 67 protein sequences. We also used about 19,000 human genes that we can confidently assume that these are not classified as antibiotic resistant. 

\begin{figure}
    \centering
    \begin{subfigure}[b]{0.4\textwidth}
        \includegraphics[width=\textwidth]{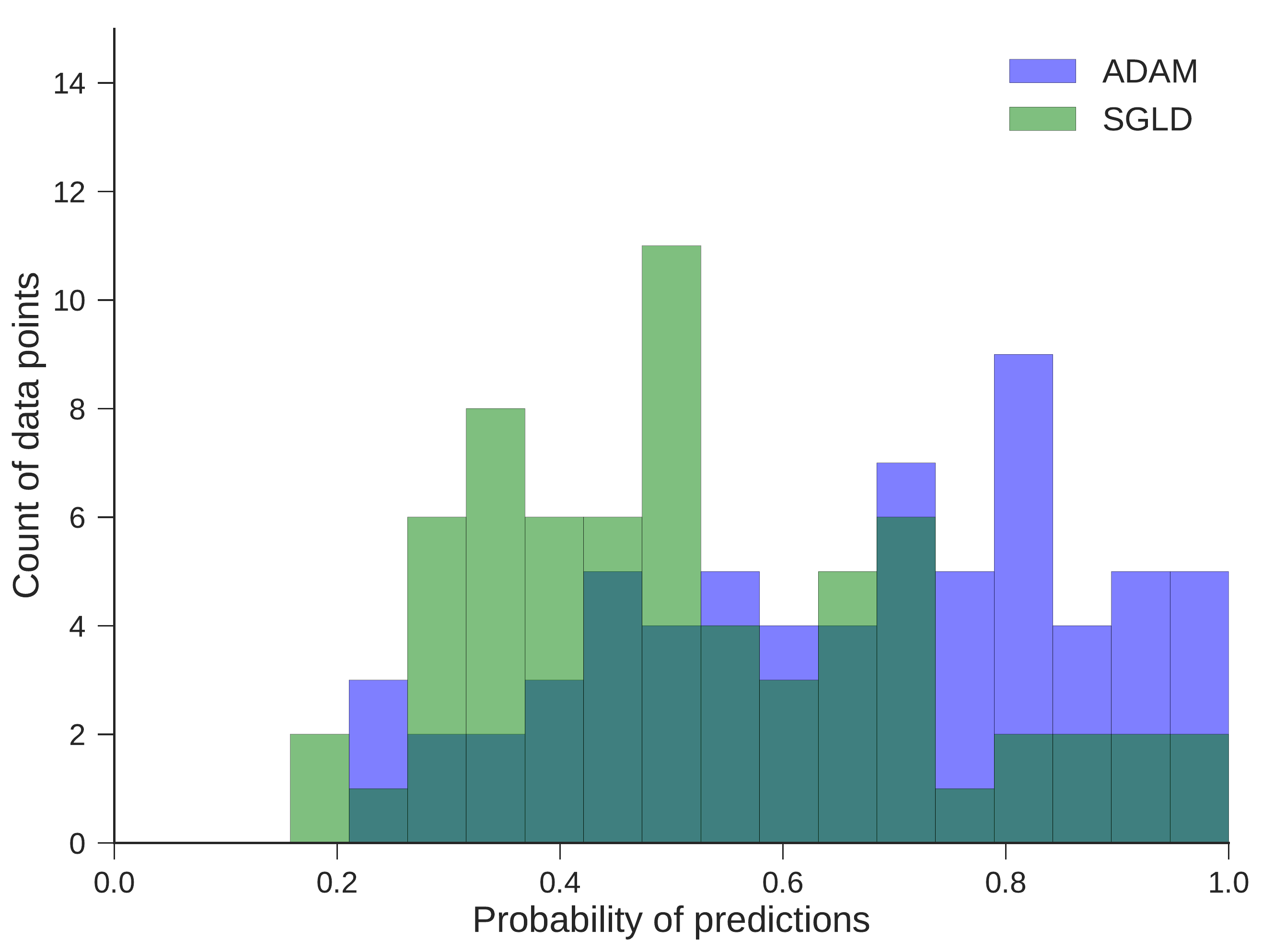}
        \caption{Probability of predictions on the 67 protein sequences that are from antibiotic resistant classes the models were not trained on}
        \label{fig:67_data}
    \end{subfigure}
    ~ 
    \hfill  
    \begin{subfigure}[b]{0.4\textwidth}
        \includegraphics[width=\textwidth]{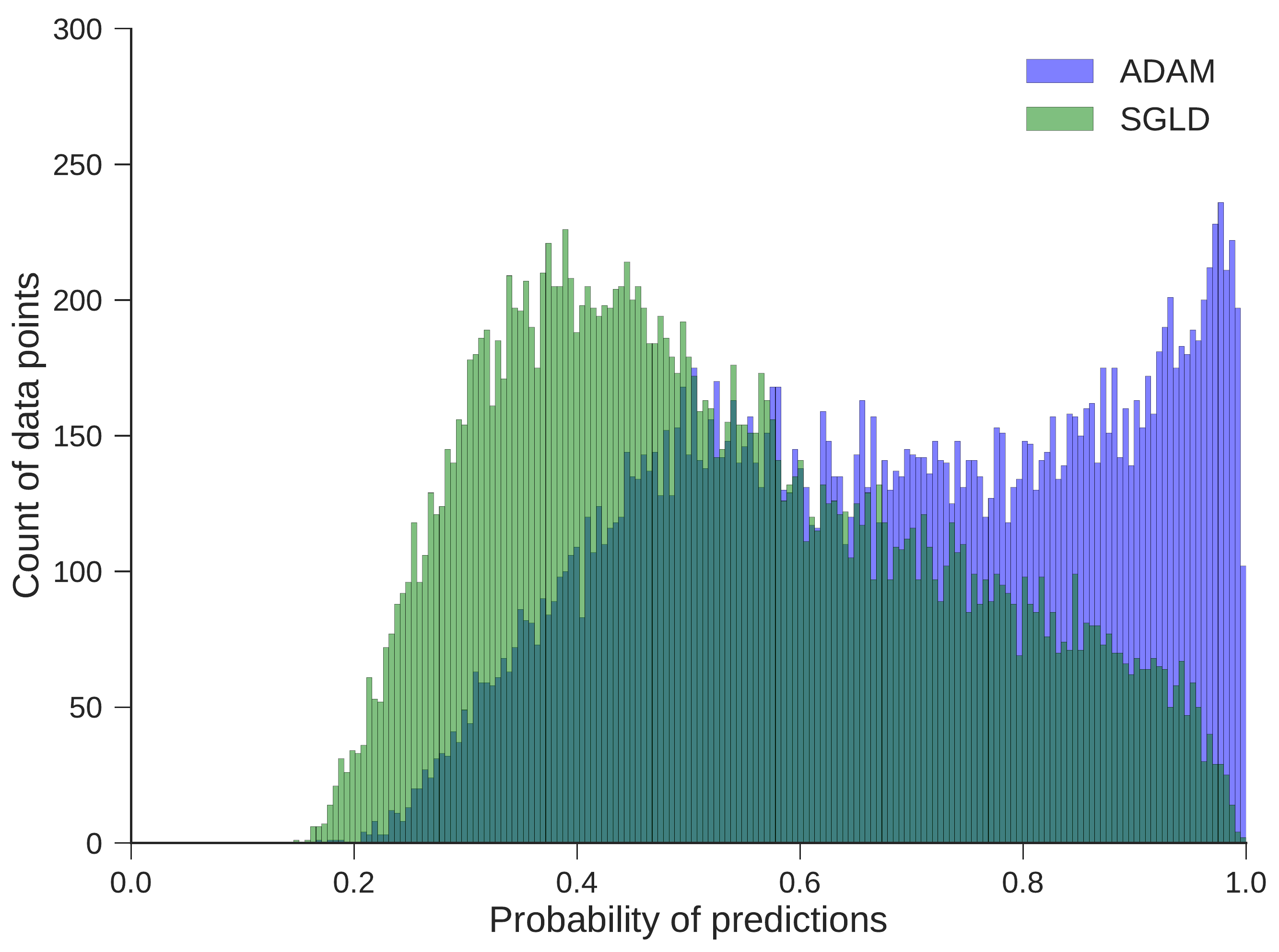}
        \caption{Approximately 19,000 human genes both models did predictions upon. These are not antibiotic resistant genes. SGLD trained neural networks predict antibiotic resistance with a much lower probability.}
        \label{fig:human_genes}
    \end{subfigure}
    \caption{\footnotesize{Probabilities assigned to predictions by both SGLD and ADAM trained models. The SGLD trained method predicts low probability for antibiotic resistance, both for classes not trained on (a) , and for genes not associated with antibiotic resistance (b).}} \label{prob_dist}
\end{figure}

Before testing on these sequences, our expectation is that an ideal model trained on a different set of classes (the 15 antibiotic classes used in training and testing in our case) should provide low probabilities for its prediction on out-of-class sequences. The model should also provide low probabilities for human genes that are not antibiotic resistant. Figure \ref{prob_dist} shows the distribution of probability scores of predictions for both the SGLD and ADAM trained model on these sequences. We can see from the figure that in both cases the probability distribution for SGLD is centered around 0.5 whereas for ADAM the distribution is heavily right skewed. The ADAM trained model is still predicting these OoD sequences to be one of the 15 classes it was trained on with high confidence. In contrast, the SGLD model is conveying its uncertainty over its predictions. 

\section{Discussion}
\label{discuss}
In this study we applied a training optimization method for neural networks which calibrates the prediction probability scores such that OoD samples are assigned low probabilities. We used this SGLD trained neural network for a multi-class classification task of antibiotic resistance type classification from protein sequences. We trained our neural network on 15 classes of antibiotic resistant proteins. We also trained another ADAM trained neural network on these same 15 classes of antibiotic resistant proteins. The overall $F_1$ score for the ADAM trained model (96\%) was higher than the SGLD trained model (91\%) model. Yet, when we tested both neural networks on two datasets of protein sequences that we know either belong to classes of antibiotic resistance that were not part of our training and testing or are not antibiotic resistance associated, the ADAM trained model still predicted them to be of the 15 classes with a high probability distribution. In contrast, for the SGLD trained model provided predictions with a lower probability distribution for the proteins not associated with antibiotic resistance. We hypothesize that the Gaussian Noise introduced in the SGLD training scheme impedes the neural networks to completely collapse on the Maximum Likelihood solution. That may also be the reason that training a neural network with SGLD towards convergence is difficult when compared with a neural network trained with ADAM and weight decay. However, SGLD lets a discriminative model detect OoD data points, and consequently provide lower probabilities in its predictions for them. This is an important property, especially when we consider the open world problem in biology where for any classification task it is hard to collect negative training samples for training the machine learning algorithm \cite{Dessimoz2013CAFA}. One avenue of future research is to investigate how to increase the accuracy of SGLD like training optimization methods. This might involve changing the structure of the noise we are introducing. 
\section{Funding}
The research is based upon work supported, in part, by the Office of the Director of National Intelligence (ODNI), Intelligence Advanced Research Projects Activity (IARPA), via the Army Research Office (ARO) under cooperative Agreement Number W911NF-17-2-0105, and by the National Science Foundation (NSF) grant ABI-1458359. The views and conclusions contained herein are those of the authors and should not be interpreted as necessarily representing the official policies or endorsements, either expressed or implied, of the ODNI, IARPA, ARO, NSF, or the U.S. Government. The U.S. Government is authorized to reproduce and distribute reprints for Governmental purposes notwithstanding any copyright annotation thereon. 

\bibliography{report}   
\bibliographystyle{unsrt}

\end{document}